\documentclass[aps,prl,reprint,superscriptaddress,nofootinbib]{revtex4-2}

\usepackage{amsmath}
\usepackage{amssymb}
\usepackage{graphicx}
\usepackage{bm}
\usepackage{hyperref}
\usepackage{xcolor}

\newcommand{\shorten}[1]{\if 0 {#1} \fi}

\newcommand{\prlsection}[1]{\par\indent\textit{#1.---}\ignorespaces}
\newcommand{\prlsubsection}[1]{\par\indent#1:\quad\ignorespaces}

\begin{document}

\title{Probing nuclear deformation through global $\Lambda$ polarization in U$+$U collisions at HIAF}

\author{Shin-ei Fujii}
\affiliation{Zhejiang Institute of Modern Physics, Department of Physics, Zhejiang University, Hangzhou 310027, China}
\affiliation{Department of Physics, Sophia University, Tokyo 102-8554, Japan}

\author{Koichi Hattori}
\affiliation{Zhejiang Institute of Modern Physics, Department of Physics, Zhejiang University, Hangzhou 310027, China}
\affiliation{Research Center for Nuclear Physics (RCNP), Osaka University, Osaka, 567-0047, Japan}

\date{\today}

\begin{abstract}
We investigate the effects of nuclear deformation on global $\Lambda$ polarization in U+U collisions at $\sqrt{s_{NN}}=2.85$~GeV using the JAM2 transport model, 
in view of future measurements at HIAF.
The tip--tip and side--side configurations yield 
substantially larger global $\Lambda$ polarization than that in the body--body configuration.
This separation grows with an increasing quadrupole deformation parameter $\beta_2$, highlighting the sensitivity of orientation-dependent polarization to nuclear deformation.
These findings connect nuclear shape with hyperon spin, opening a new avenue for probing nuclear deformation through polarization measurements in intermediate-energy heavy-ion collisions.
\end{abstract}

\maketitle

\prlsection{Introduction}
Spin polarization has emerged as a powerful probe of the rotational dynamics of matter created in heavy-ion collisions.
In non-central collisions, a large orbital angular momentum is generated perpendicular to the reaction plane, 
creating a vortical structure of the produced matter \cite{Liang:2004ph}. 
Such vorticity can be transferred to the spin degrees of freedom of produced particles 
through spin--vorticity coupling~\cite{Becattini:2013fla, Becattini:2016gvu}, leading to a global polarization of hadrons along the angular momentum (see, e.g., Refs.~\cite{Becattini:2020ngo, Huang:2020dtn, Becattini:2024uha,Niida:2024ntm} for reviews).
The first statistically significant observation of global $\Lambda$ and $\bar{\Lambda}$ polarization by the STAR Collaboration established spin polarization as a new observable for exploring the dynamical properties of strongly interacting matter~\cite{STAR:2017ckg}. 

Of particular interest is the intermediate-energy region of a few GeV.
Measurements by the STAR Collaboration in fixed-target Au+Au collisions at $\sqrt{s_{NN}}=3$~GeV revealed a sizable global $\Lambda$ polarization~\cite{STAR:2021beb}.
At even lower energies, the HADES Collaboration measured large $\Lambda$ polarization in Au+Au and Ag+Ag collisions at $\sqrt{s_{NN}}=2.4$ and $2.55$~GeV, respectively~\cite{HADES:2022enx}.
These observations extend the increasing trend of global polarization toward lower collision energies and indicate that the few-GeV region provides a particularly interesting environment for studying spin dynamics in strongly interacting matter.
Recent theoretical calculations also suggest that the global $\Lambda$ polarization may exhibit a broad maximum around $\sqrt{s_{NN}}\sim3$--$4$~GeV~\cite{Ivanov:2022ble, Ivanov:2025izv}.

The High Intensity heavy-ion Accelerator Facility (HIAF)~\cite{Zhou:2022pxl}, which has recently entered trial operation in China, will provide us with new opportunities to explore heavy-ion collisions in this intermediate-energy and high-baryon-density regime.
The proposed Hyperon-Nucleon Spectrometer (H-NS) at HIAF aims to perform systematic measurements of hyperon polarization in $pp$, $pA$, and $AA$ collisions, opening prospects for dedicated studies of spin phenomena in this energy range~\cite{Bai:2026syt}.

A particularly intriguing opportunity at HIAF is provided by collisions between strongly deformed nuclei such as $^{238}$U.
The intrinsic shape of the colliding nuclei has recently attracted considerable attention 
at higher collision energies.
At the RHIC energies, the 
nuclear crossing time, of the order of $0.1$~fm/$c$, is much shorter than the characteristic timescale of rotational quantum fluctuations of a deformed $^{238}$U nucleus, $\tau_{\rm rot}\sim10^3$--$10^4$~fm/$c$~\cite{STAR:2024wgy}.
Consequently, the nuclear orientation is effectively frozen during the collision, allowing each collision event to capture a ``snapshot'' of the intrinsic nuclear shape.
This idea has enabled 
the nuclear-shape imaging using high-energy U+U collisions at RHIC~\cite{STAR:2024wgy, Jia:2025wey, STAR:2025elk}.

The situation at intermediate collision energies is qualitatively different.
The nuclear crossing time increases to several fm/$c$ which remains orders of magnitude shorter than the rotational timescale of $^{238}$U.
The orientation of the deformed nucleus can therefore still be regarded as essentially frozen throughout the collision.
However, unlike at ultrarelativistic energies, where the nuclear geometry is rapidly encoded into the initial condition of the produced medium, the finite crossing time at intermediate energies allows nuclear deformation to influence the collision dynamics throughout the overlap stage. 
Consequently, the intrinsic nuclear shape can affect compression, baryon stopping, and rescattering, thereby modifying the subsequent vortical evolution of the system.

This raises a fundamental question: Can the intrinsic shape of colliding nuclei leave a characteristic imprint on the spin polarization of produced hadrons through the dynamical generation of vorticity?
Different orientations of the prolate $^{238}$U nuclei modify not only the collision geometry but also the distribution of orbital angular momentum and the resulting vortical structure of the system.
Through spin--vorticity coupling, these orientation-dependent vortical structures are expected to leave characteristic imprints on the global polarization of produced hadrons.
Therefore, spin polarization in the intermediate collision energies is not only of interest by its own but also may 
serve as 
a novel probe of nuclear deformation. 
\shorten{, 
complementary to established deformation-sensitive observables at high energies, such as particle multiplicity and anisotropic flow.
}

In this Letter, we 
pursue this possibility using the Jet AA Microscopic Transport Model (JAM2)~\cite{Nara:2021fuu, Sahoo:2023fzz, Nara:2025pkg}.
We simulate U+U collisions at $\sqrt{s_{NN}}=2.85$~GeV, relevant to the HIAF energy regime, and systematically study the dependence of the vorticity and global $\Lambda$ polarization on the orientation and deformation of the colliding uranium nuclei.
We also compare our calculations for Au+Au collisions at $\sqrt{s_{NN}}=3$~GeV with the STAR measurements to assess the framework in a nearby energy regime.
Our aim is to determine how nuclear-shape information is reflected in global $\Lambda$ polarization and to provide predictions for future measurements at HIAF.

Throughout this study, we employ natural units, $\hbar = c = k_B = 1$, and use the Minkowski metric $g_{\mu\nu} = \mathrm{diag}(+,-,-,-)$ and the Levi-Civita tensor with the convention $\epsilon^{0123}=+1$.

\prlsection{Polarization from the transport model}
To evaluate the spin polarization within a transport model, we first reconstruct the vortical structure of the medium from the microscopic particle distributions. 
Since transport models describe the system in terms of individual particles, the local fluid velocity and temperature fields required for evaluating the thermal vorticity 
\shorten{are not directly available and} 
need to be reconstructed through a coarse-graining procedure~\cite{Huang:2020dtn}.

\prlsubsection{Vorticity}
We define the thermal vorticity as
\begin{equation}
\varpi_{\mu\nu}
=
-\frac{1}{2}
(\partial_\mu\beta_\nu-\partial_\nu\beta_\mu),
\end{equation}
where $\beta^\mu = u^\mu/T$ with the fluid four-velocity $u^\mu$ and temperature $T$.
The local rotational structure is characterized by the thermal-vorticity pseudovector 
$\varpi^\mu = -\frac{1}{2} \epsilon^{\mu\nu\rho\sigma} u_\nu \varpi_{\rho\sigma}$,
which satisfies $u_\mu\varpi^\mu=0$ and is thus purely spatial in the local rest frame.

The coarse-grained energy--momentum tensor and baryon current are constructed from the particle distributions as
\begin{align}
T^{\mu\nu}(x) 
&= \sum_i \frac{p^\mu_i p^\nu_i}{p^0_i} 
W_i(\bm{x}),\\
J^\mu_B(x)
&= \sum_i B_i \frac{p_i^\mu}{p_i^0}
W_i(\bm{x}),
\end{align}
where $p_i^\mu$ and $B_i$ are the four-momentum and baryon number of particle $i$, respectively. 
We employ a normalized Gaussian smearing 
\begin{equation}
W_i(\bm{x})
=
\frac{1}{(2\pi\sigma^2)^{3/2}}
\exp\left[
-\frac{|\bm{x}-\bm{x}_i|^2}{2\sigma^2}
\right],
\end{equation}
with the particle position $\bm{x}_i$ 
and the smearing length $\sigma = 1~\mathrm{fm}$.
The fluid four-velocity is defined in the Eckart frame by the baryon current,
\begin{equation}
u^\mu(x)
=
\frac{J_B^\mu(x)}
{\sqrt{J_B^\nu(x)J_{B\nu}(x)}}.
\end{equation}
The energy density and baryon-number density are obtained as 
$e=u_\mu T^{\mu\nu}u_\nu
$ and $n_B=u_\mu J_B^\mu$, respectively.

We determine the temperature entering the thermal-vorticity calculation by matching these local densities to a hadron resonance gas (HRG) equation of state, yielding $T=T(e,n_B)$ with assumptions $n_Q = \frac{92}{238}n_B$ and $n_S = 0$.
The hadron species included in the HRG are chosen to match those implemented in JAM2, ensuring consistency between the microscopic particle content and the thermodynamic description.
The use of the HRG equation of state is further motivated by recent SMASH~\cite{SMASH:2016zqf} calculations, in which an HRG-based determination of the temperature yields good agreement with measurements of $\Lambda$ polarization at intermediate collision energies, including Au+Au collisions at $\sqrt{s_{NN}}=2.4$~GeV and $\sqrt{s_{NN}}=3$~GeV~\cite{Yi:2026rbz}.

\prlsubsection{Polarization}
Once the thermal vorticity is determined, the mean spin four-vector of a spin-$1/2$ particle in a local thermal equilibrium is given by~\cite{Becattini:2013fla, Becattini:2016gvu}
\begin{equation}
S^\mu(x,p) =
-\frac{1}{8m}
\left[1-n_{\mathrm F}(x,p)\right]
\epsilon^{\mu\nu\rho\sigma}
p_\sigma
\varpi_{\nu\rho}(x),
\end{equation}
where $m$ and $n_{\mathrm F}$ denote the particle mass and the Fermi--Dirac distribution function, respectively.
Since the $\Lambda$ mass is much larger than typical freeze-out temperatures, we treat $\Lambda$ hyperons in the dilute Boltzmann limit and neglect the Pauli-blocking correction, setting $1-n_{\mathrm F}(x,p)\simeq 1$~\cite{Becattini:2013vja}.
Following the transport-model prescription of Ref.~\cite{Vitiuk:2019rfv}, we identify the freeze-out point of each $\Lambda$ hyperon with its last interaction point, $x_{\mathrm{last}}$, and evaluate its mean spin four-vector using the local thermal vorticity at that point.

The polarization vector $\bm{P}$ is defined with the spatial component $\bm{S}^{*}$ of the spin four-vector in the particle rest frame as 
\begin{equation}
\bm{P}=2\bm{S}^{*}, \qquad 
\bm{S}^{*}
=
\bm{S}
-
\frac{\bm{p}\cdot\bm{S}}
{E(m+E)}
\bm{p} ,
\end{equation}
where $E$ and $\bm{p}$ are the particle energy and momentum in the computational frame, respectively. 
It is obtained by Lorentz transforming the spin four-vector from the computational frame to the particle rest frame. 
The global polarization is obtained by averaging the rest-frame spin polarization along the direction of the total orbital angular momentum over the selected $\Lambda$ hyperons.
\shorten{[The last two paragraphs merged.]}


\prlsection{The JAM2 model}
In this study, we employ the Jet AA Microscopic Transport Model (JAM2)~\cite{Nara:2021fuu, Sahoo:2023fzz, Nara:2025pkg}, which describes the dynamical evolution of hadrons in heavy-ion collisions over a wide range of collision energies.

In JAM2, the nuclear deformation of $^{238}\mathrm{U}$ is implemented using a deformed Woods--Saxon distribution~\cite{Hagino:2006fj},
\begin{align}
\rho(r,\theta,\phi)
&=
\frac{\rho_0}
{1+\exp\left[(r-R(\theta,\phi))/a_0\right]},
\\
R(\theta,\phi)
&=
R_0
\left[
1+\beta_2Y_2^0(\theta,\phi)
+\beta_4Y_4^0(\theta,\phi)
\right],
\end{align}
where $\rho_0$, $R_0$, and $a_0$ denote the central nuclear density, nuclear radius, and surface diffuseness, respectively. The deformation parameters $\beta_2$ and $\beta_4$, together with the spherical harmonics $Y_l^m(\theta,\phi)$, characterize the quadrupole and hexadecapole deformations of the uranium nucleus, respectively. 
We use $\rho_0=0.166~\mathrm{fm}^{-3}$, $R_0=6.86~\mathrm{fm}$, $a_0=0.44~\mathrm{fm}$, $\beta_2=0.28$, and $\beta_4=0.093$ as default parameters.

\begin{figure}[t]
\includegraphics[width=0.95\linewidth]{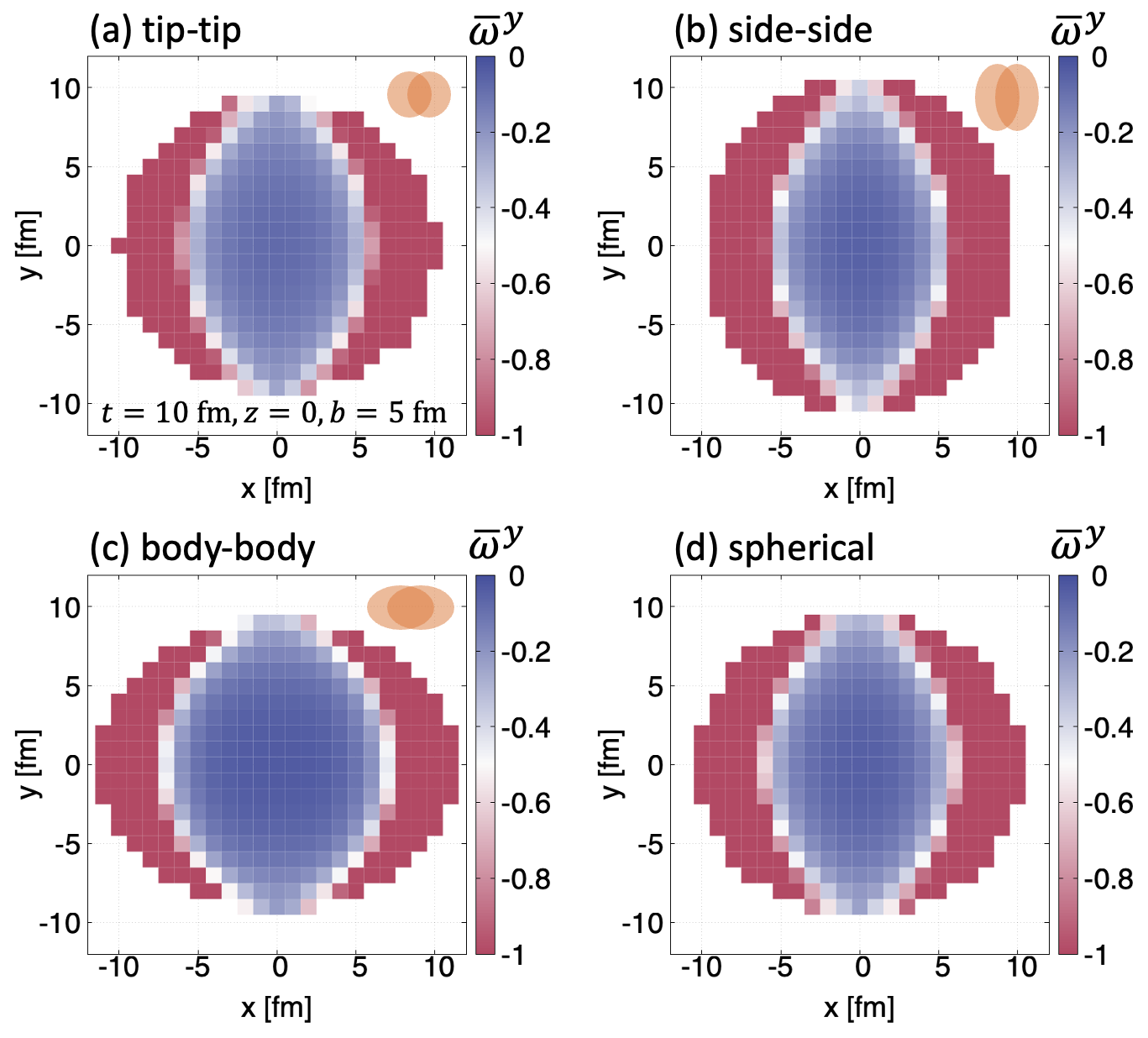}
\caption{Spatial distributions of the $y$ component of the averaged thermal-vorticity pseudovector, $\varpi^{y}$, in the transverse ($x$--$y$) plane at $z=0$ and $t=10$~fm for an impact parameter $b=5$~fm.
Panels (a)--(d) correspond to tip--tip, side--side, body--body, and spherical configurations, respectively.  
}
\label{fig:vorticity}
\end{figure}

In JAM2, the orientation of each deformed uranium nucleus can be specified with respect to the collision geometry. 
The tip--tip, body--body, and side--side configurations correspond to the symmetry axes of both prolate $^{238}\mathrm{U}$ nuclei being aligned along the beam direction $z$, 
the impact-parameter direction $x$, and 
the remaining out-of-plane direction $y$, respectively (cf. Fig.~\ref{fig:vorticity}).

JAM2 provides several options for the mean-field interaction.
In this study, we use the RQMDv mean field, in which a momentum-dependent Skyrme-type interaction is implemented as a Lorentz four-vector potential.
The momentum-dependent soft parameter set MS2, corresponding to a nuclear incompressibility of $K=210~\mathrm{MeV}$, is adopted (see Ref.~\cite{Nara:2021fuu} for details).

\prlsection{Results}
In the results section, we focus on the global polarization along the $y$ direction, perpendicular to the reaction plane, and denote the global $\Lambda$ polarization by $P_{\Lambda}$.
Throughout this section, error bars for the JAM2 polarization results represent standard errors of the event-averaged polarization.

Figure~\ref{fig:vorticity} shows the spatial distributions of the $y$ component of the averaged thermal vorticity in the transverse plane.
In all four configurations, regions of large vorticity magnitude are concentrated near the outer edges of the distribution, surrounding a central region of relatively weak vorticity.
The body--body configuration exhibits the most distinct spatial pattern, with the weak-vorticity region occupying a larger fraction of the distribution than in the other configurations.
Since the $y$-directed vorticity is driven predominantly by the longitudinal shear across the impact-parameter direction, this
behavior can be understood from the broader extent of the body--body overlap region along $x$. 
That is, the longitudinal-flow variation is distributed over a larger transverse distance, reducing the local velocity gradient
and thereby the global $\Lambda$ polarization as discussed below.

\begin{figure*}[t]
\centering 
\includegraphics[width=0.95\textwidth]{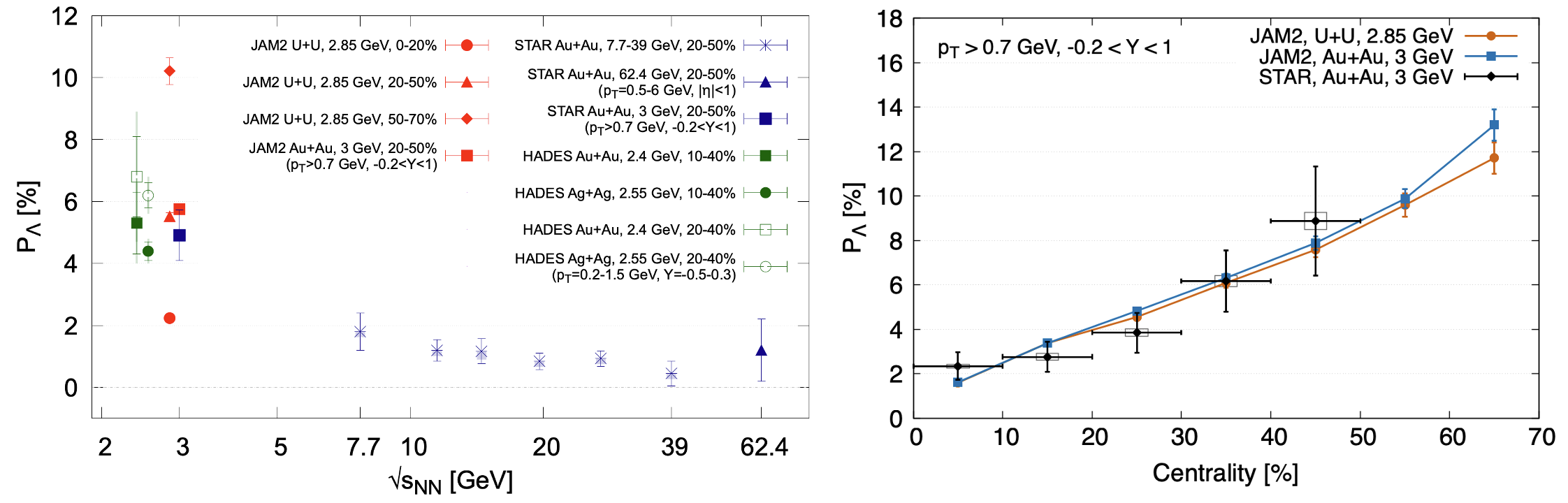} 
\caption{Global $\Lambda$ polarization as a function of collision
energy (left) and centrality (right).
In the left panel, the JAM2 results are shown for U+U collisions at
$\sqrt{s_{NN}}=2.85$~GeV in the 0--20\%, 20--50\%, and 50--70\%
centrality classes, and for Au+Au collisions at
$\sqrt{s_{NN}}=3$~GeV in the 20--50\% centrality class.
Also shown are STAR measurements for Au+Au collisions at
$\sqrt{s_{NN}}=3$~GeV~\cite{STAR:2021beb},
7.7--39~GeV~\cite{STAR:2017ckg}, and
62.4~GeV~\cite{STAR:2007ccu}, together with HADES measurements for
Au+Au and Ag+Ag collisions at $\sqrt{s_{NN}}=2.4$ and
2.55~GeV, respectively~\cite{HADES:2022enx}.
The right panel compares the centrality dependence obtained from JAM2
for U+U collisions at 2.85~GeV and Au+Au collisions at 3~GeV with the
STAR Au+Au data at 3~GeV.
All JAM2 results are evaluated using the STAR kinematic cuts,
$p_T>0.7$~GeV and $-0.2<Y<1$.}
\label{fig:polarization_energy}
\end{figure*}

Figure~\ref{fig:polarization_energy} (left) shows the collision-energy dependence of global $\Lambda$ polarization, with the JAM2 results indicated by red markers. 
For U+U collisions at $\sqrt{s_{NN}}=2.85$~GeV, the predicted polarization increases toward peripheral collisions, from $(1.963\pm0.026)\%$ in the 0--20\% centrality class to $(4.922\pm0.045)\%$ and $(9.59\pm0.15)\%$ in the 20--50\% and 50--70\% classes, respectively.
For Au+Au collisions at $\sqrt{s_{NN}}=3$~GeV, the calculation yields $(5.752\pm0.097)\%$ in the 20--50\% centrality class, consistent with the STAR measurement within the experimental uncertainties~\cite{STAR:2021beb}.

Figure~\ref{fig:polarization_energy} (right) shows the centrality dependence of the global $\Lambda$ polarization calculated with JAM2.
The results for U+U collisions at $\sqrt{s_{NN}}=2.85$~GeV and Au+Au collisions at $\sqrt{s_{NN}}=3$~GeV exhibit similar centrality dependence, with slightly smaller polarization values for U+U collisions across the centrality range shown.
For Au+Au collisions, the calculation remarkably reproduces the STAR measurements~\cite{STAR:2021beb}.
This agreement provides an empirical benchmark for the polarization calculation at an energy close to that considered here and supports its application to U+U collisions in the HIAF energy regime.

Building on the agreement with the STAR measurements demonstrated in Fig.~\ref{fig:polarization_energy}, we now examine how nuclear orientation and deformation affect global $\Lambda$ polarization in U+U collisions at $\sqrt{s_{NN}}=2.85$~GeV.

\begin{figure*}[t]
\includegraphics[width=0.95\textwidth]{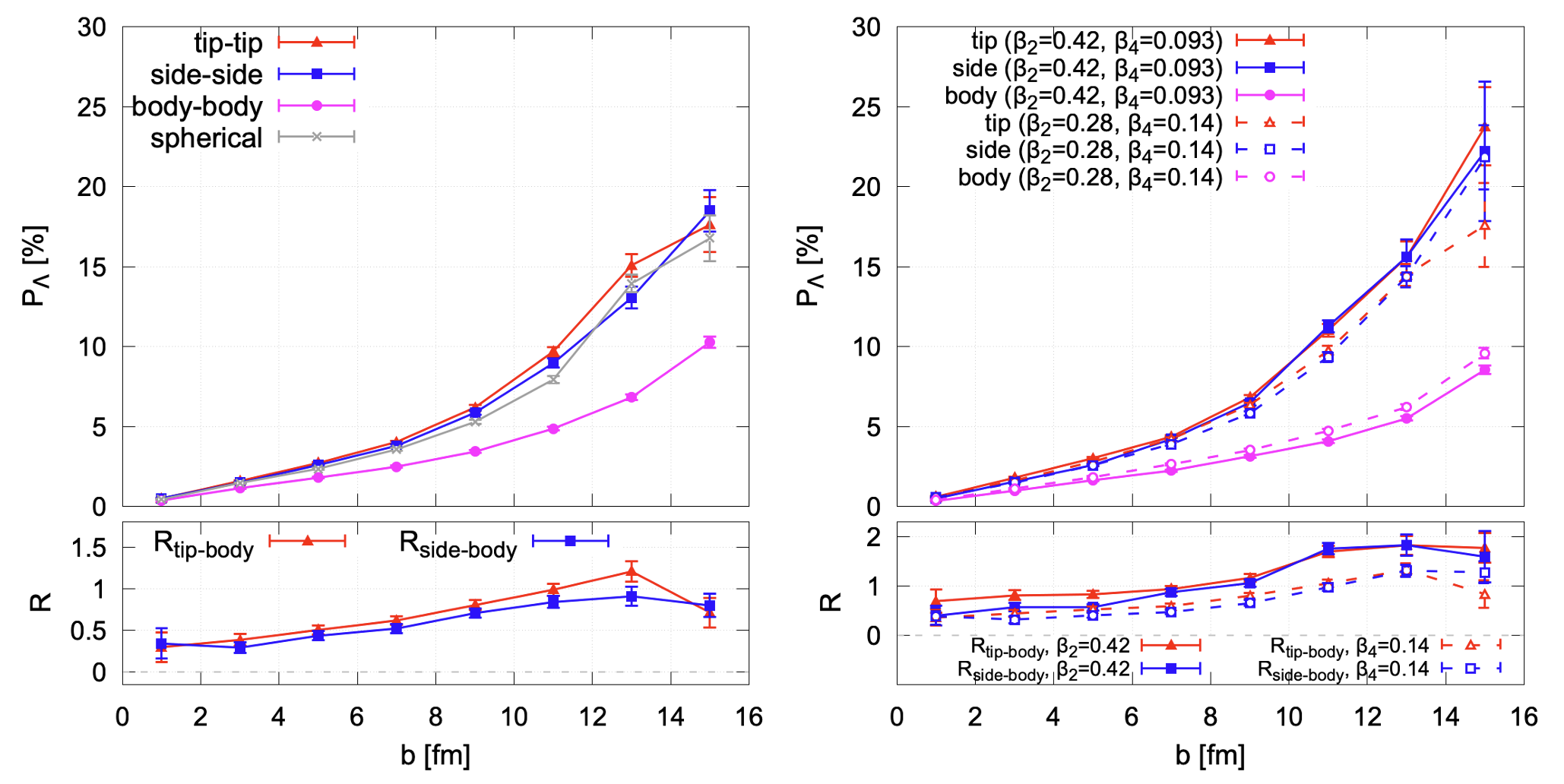}
\caption{Orientation and deformation dependence of the global
$\Lambda$ polarization in U+U collisions at
$\sqrt{s_{NN}}=2.85$~GeV as a function of impact parameter.
The upper-left panel compares the polarization for the tip--tip,
side--side, body--body, and spherical configurations using the default 
deformation parameters.
The upper-right panel shows the corresponding results for the tip--tip,
side--side, and body--body configurations with
$(\beta_2,\beta_4)=(0.42,0.093)$ (solid lines and filled symbols) and
$(0.28,0.14)$ (dashed lines and open symbols).
The lower panels show the relative polarization differences
$R_{\mathrm{tip-body}}$ and $R_{\mathrm{side-body}}$, which quantify
the deviations of the tip--tip and side--side results, respectively,
from the body--body result, as defined in the text.}
\label{fig:polarization_b}
\end{figure*}

The upper panel of Fig.~\ref{fig:polarization_b} (left) shows the global $\Lambda$ polarization as a function of impact parameter for different nuclear orientations.
The tip--tip, side--side, and spherical configurations yield comparable polarization magnitudes, with the tip--tip results being slightly larger than the side--side results, followed by the spherical results.
In contrast, the body--body configuration exhibits distinctly smaller polarization, producing a pronounced separation from the other three configurations.

To quantify the orientation dependence, we define the relative
polarization difference with respect to the body--body configuration as
\begin{equation}
R_{X-\mathrm{body}}
\equiv
\frac{
P_{\Lambda,X}-P_{\Lambda,\mathrm{body}}
}{
P_{\Lambda,\mathrm{body}}
},
\quad
X\in\{\mathrm{tip},\mathrm{side}\}.
\end{equation}
The lower panel of Fig.~\ref{fig:polarization_b} (left) shows the resulting $R_{\mathrm{tip-body}}$ and $R_{\mathrm{side-body}}$.
Both generally increase toward peripheral collisions, with $R_{\mathrm{tip-body}}$ slightly exceeding $R_{\mathrm{side-body}}$.
The magnitude of these relative differences and their implications for probing nuclear deformation are discussed in detail below in Fig.~\ref{fig:R}.

To assess 
the sensitivity of global $\Lambda$ polarization to nuclear deformation, 
we increase each of $\beta_2$ and $\beta_4$ by 50\% in separate calculations
while keeping the other deformation parameter at its default value.
The upper panel of Fig.~\ref{fig:polarization_b} (right) clearly shows that increasing $\beta_2$ from $0.28$ to $0.42$ enhances the separation of the tip--tip and side--side results from the body--body results.
In comparison, increasing $\beta_4$ from $0.093$ to $0.14$ produces only modest changes. 

The lower panel of Fig.~\ref{fig:polarization_b} (right) shows that this parameter dependence is also reflected in the relative polarization differences.
Increasing $\beta_2$ (solid lines) enhances both $R_{\mathrm{tip-body}}$ and $R_{\mathrm{side-body}}$, whereas the results obtained with enhanced $\beta_4$ (dashed lines) remain close to those for the default parameters.

To compare the deformation dependence more directly, Fig.~\ref{fig:R} brings together $R_{\mathrm{tip-body}}$ for the default and modified deformation parameter sets shown in Fig.~\ref{fig:polarization_b}.
Since $R_{\mathrm{tip-body}}$ vanishes for spherical nuclei, its nonzero value directly reflects nuclear deformation, making it a particularly useful observable for probing deformation parameters within the present framework.
The relative difference generally grows toward peripheral collisions. 
\shorten{, although it decreases at the largest impact parameter shown.} 
For the default deformation parameters, it rises from approximately $0.3$ at small impact parameters to above unity in peripheral collisions.
Increasing $\beta_2$ to $0.42$ further enhances this contrast: $R_{\mathrm{tip-body}}$ varies from $0.7$ to maximally reach $1.8$ approximately. 
These clear tendency suggest that 
nuclear-shape information can be reflected in the hyperon spin polarization. 
By comparison, increasing $\beta_4$ produces only a modest enhancement relative to the default results. 
\shorten{
The large relative differences and their pronounced response to $\beta_2$ show that nuclear-shape information can be reflected in hyperon spin polarization.
Orientation-dependent polarization measurements may therefore provide a complementary means of constraining nuclear quadrupole deformation.
}
\begin{figure}[t]
\includegraphics[width=0.95\linewidth]{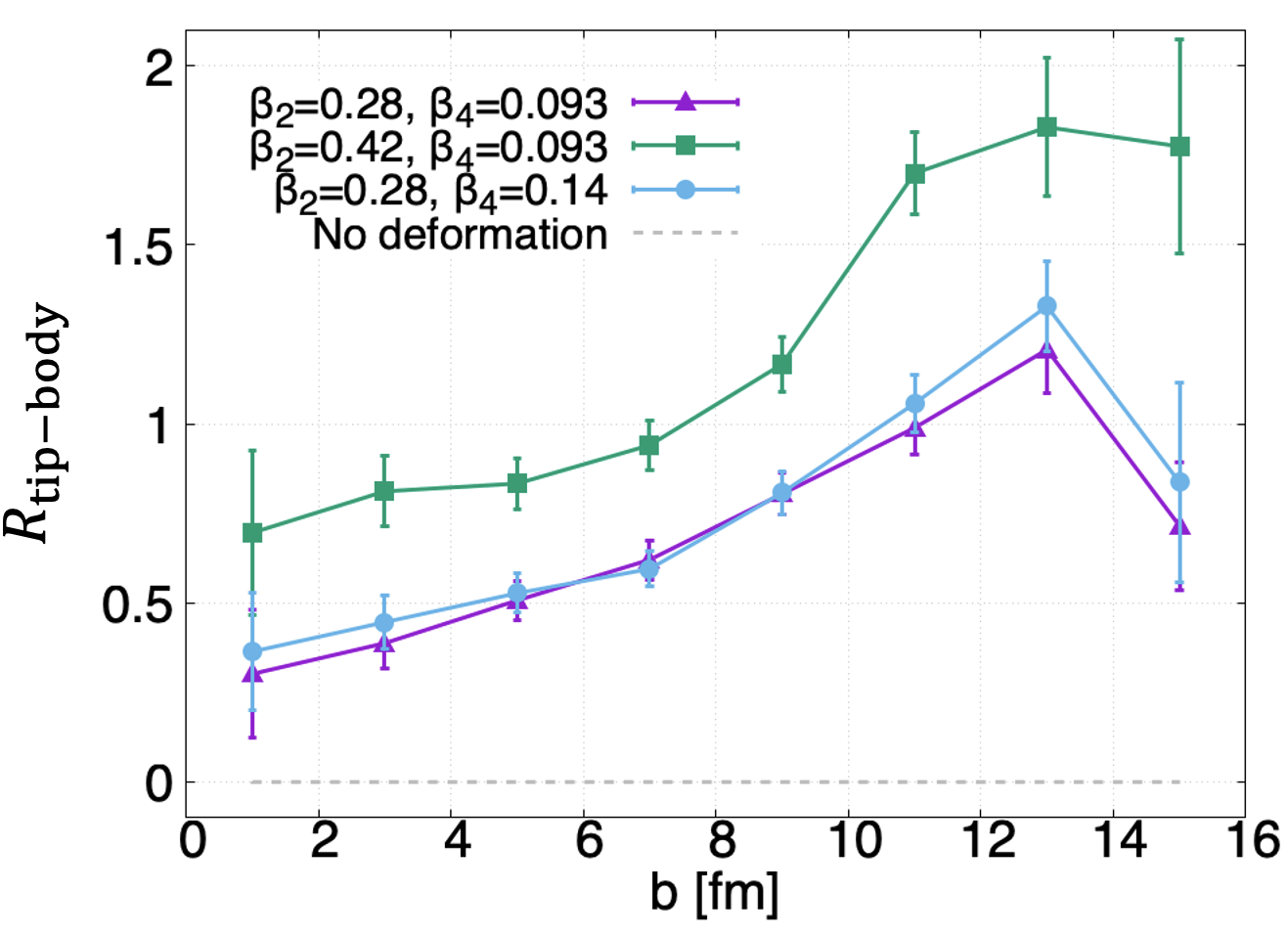}
\caption{Relative polarization difference between tip--tip and body--body configurations, $R_{\mathrm{tip-body}}$, as a function of impact parameter.
The curves show results for the default deformation parameters, $(\beta_2,\beta_4)=(0.28,0.093)$ (purple), enhanced quadrupole deformation, $(0.42,0.093)$ (green), and enhanced hexadecapole deformation, $(0.28,0.14)$ (light blue).
For spherical nuclei, the two configurations are equivalent, and $R_{\mathrm{tip-body}}$ is expected to vanish (dashed gray).}
\label{fig:R}
\end{figure}

\prlsection{Summary}
In this Letter, we investigated the connection between nuclear deformation and global $\Lambda$ polarization in U+U collisions at $\sqrt{s_{NN}}=2.85$~GeV using the JAM2 transport model to provide predictions for future experiments at HIAF.
The calculations reproduce the STAR polarization measurements in Au+Au collisions at $\sqrt{s_{NN}}=3$~GeV within the experimental uncertainties~\cite{STAR:2021beb}, providing an empirical benchmark for predictions in the nearby HIAF energy regime.

This agreement motivates us to further investigate
how nuclear geometry influences the spin polarization. 
Global $\Lambda$ polarization is substantially smaller in body--body collisions than in tip--tip and side--side collisions.
This contrast becomes more pronounced with increasing $\beta_2$, while the response to $\beta_4$ is modest.
The relative tip--body difference, $R_{\mathrm{tip-body}}$, exceeds unity in peripheral collisions for the default parameters and reaches approximately $1.8$ for $\beta_2=0.42$, highlighting its sensitivity to quadrupole deformation. 

These results reveal how the intrinsic shape of colliding nuclei leaves a distinct imprint on hyperon spin polarization through the dynamical generation of vorticity, establishing a new connection between nuclear structure and spin dynamics in intermediate-energy heavy-ion collisions.
Beyond the global polarization, this new approach may shed light on other spin phenomena, including local hyperon polarization and vector-meson spin alignment.

\prlsection{Acknowledgments}
We are grateful to Y.~Nara for valuable discussions on the JAM2 transport model used in this study.
This work is supported in part by JSPS KAKENHI Grant Number~23K22487 and National Science Foundation of China under grant No.~W2532002. 

\bibliographystyle{apsrev4-2}
\bibliography{reference}

\end{document}